\documentclass[intlimits,twoside,a4paper]{article}

\usepackage{graphicx}

\usepackage[eqsecnum]{cmpj3}

\usepackage{bm}

\issue{2026}{29}{3}{33702}
\doinumber{10.5488/CMP.29.33702}

\title[Tailoring structural, electronic, optical, and magnetic properties]
{Tailoring structural, electronic, optical, and magnetic properties of rare-earth gallates RGaO$_3$ (R = Ho, Er, Tm) via first-principles investigations}

\author[T. Usman, K. Liaqat, S. Khan, M. Yar Khan, S. Ali Khan]
{T. Usman\orcid{0000-0002-7796-425X}\refaddr{label1}{\thanks{Corresponding author: tariqusman@qlit.edu.cn}}, 
		K. Liaqat\orcid{0000-0002-8819-6903}\refaddr{label2}, S.~Khan\orcid{0009-0002-9053-0242}\refaddr{label3}, M.~Yar~Khan\orcid{0000-0002-7376-6790}\refaddr{label1}, S.~Ali~Khan\orcid{0000-0002-8819-6903}\refaddr{label4}{\thanks{Corresponding author: salman@sdu.edu.cn}}}
		
\addresses{
	\addr{label1} Department of Physics, Qilu Institute of Technology, Jinan 250200, Shandong, P.R.~China
\addr{label2} Kohat University of Science and Technology (KUST), Kohat, Pakistan
\addr{label3} Department of Physics, University of Science \& Technology Bannu, Bannu 28100, Pakistan
\addr{label4} School of Materials Science \& Engineering, Shandong University, Jinan 25000, P.~R.~China
}
\date{Received 1 October 2025; revised 23 March 2026; accepted 27 May 2026; published 28 September 2026}

\begin{document}
	
	\maketitle
	
	\begin{abstract}
Ab initio calculations of cubic RGaO$_3$ (R = Ho, Er, Tm) perovskites were performed using the FP-LAPW method within DFT. Their structural, electronic, magnetic, and optical properties were investigated. The optimized lattice constants range from 3.83 to 3.85~\AA. ErGaO$_3$ is semiconducting, whereas HoGaO$_3$ and TmGaO$_3$ show half-metallic behavior, likely due to spin--orbit coupling near the Fermi level. Optical response of half metallic compounds (HoGaO$_3$,TmGaO$_3$) is strongly governed by their spin polarization electroni structure. The semiconducting spin channel absorbs ultraviolet photons , while metallic or small gap channel interacts with infrared light. This dual behavior enables spin selective optical excitation, relevant for spintronics and optoelectronics applications. These compounds show high surface reflectivity, with TmGaO$_3$ achieving a maximum reflectivity of 69\%. Moreover, their optical conductivity mainly occurs at higher energies, indicating suitability for high-energy optoelectronic applications. The calculated magnetic moments suggest robust ferromagnetic character across this series.

\keywords rare-earth gallates, half-metallic, density functional theory, optoelectronic properties, magnetic moments, spintronic devices
\end{abstract}


\section{Introduction}
Perovskite materials have attracted great attention in modern condensed matter physics and materials chemistry due to their wide-ranging functional properties and remarkable structural versatility. More broadly, the perovskite structure can be described by the general formula ABX$_3$, where A and B denote cations of varying size and valence (often alkali metals, alkaline earth metals, transition metals, or rare-earth elements), and X is typically an oxygen or halogen anion. This typical structure can accommodate diverse chemical substitutions, enabling the synthesis of compounds that range from insulators to semiconductors and, in some cases, half-metallic ferromagnets. The ideal cubic perovskite phase, assigned to the Pm\={3}m space group, exhibits high symmetry in which the A-site cation occupies the cube corners, the B-site cation is octahedrally coordinated at the body center, and the anions are located at face-centered positions. However, subtle distortions of the BO$_6$ octahedra frequently lead to lower-symmetry variants such as tetragonal, orthorhombic, monoclinic, or triclinic phases~\cite{1,2,3,4}. These structural distortions play a crucial role in determining the electronic structure, lattice dynamics, and functional responses of the material.

Extensive experimental and theoretical study has shown that perovskites are key candidates in optoelectronic devices, scintillators, magnetic storage, and solid oxide fuel cells applications~\cite{5,6,7,8,9}. For instance, halide perovskites, in which X is a halogen (F, Cl, Br, or I), have attracted great interest in photovoltaic technologies due to their favorable band gaps and high optical absorption coefficients~\cite{10,11,12}. By contrast, inorganic oxide perovskites, including those based on rare-earth cations, are valued for their mechanical stability, radiation hardness, and magnetic ordering phenomena~\cite{13,14,15,16}. Particularly, relevant to this study are perovskites containing rare-earth elements at the A-site and trivalent cations at the B-site. In these systems, the 4f electrons of rare-earth ions carry large localized magnetic moments, leading to an intriguing coupling between lattice, spin, and orbital degrees of freedom. For example, recent investigation by Tariq et al.~\cite{17} demonstrated that the various characterization of rare-earth cations in RAlO$_3$ (R = Sm, Eu, Gd, Dy, Tb, Ho, Tm, Er, Yb) yields rich combinations of structural, elastic, electronic, and magnetic properties. Similarly, Theofylaktos et al.~\cite{18} investigate perovskite materials with s- and d-block metals, highlighting the importance of outer electron configurations in stabilizing diverse electronic phases. Initially, the orthorhombic phase of rare-earth gallates (RGaO$_3$) was synthesized in the late 1960s and revealed to adopt the GdFeO$_3$-type structure under ambient conditions~\cite{19,20}. In the 1970s, researchers succeeded in stabilizing hexagonal forms of RGaO$_3$ compounds for elements such as Gd, Tb, and Dy~\cite{21}. These materials are notable for their phosphorescence, luminescence, and potential as laser hosts~\cite{22,23}. Mixed systems in which Ga is partially substituted by Mn demonstrate further complexity, exhibiting giant magnetostriction akin to manganites~\cite{24,25}. Recent developments in computational materials science have considerably advanced the understanding of these systems. First-principles calculations based on density functional theory (DFT) have confirmed to describe the relationship between structural stability, electronic correlations and magnetic exchange. Using FP-LAPW approaches implemented in the WIEN2k code, several studies have successfully predicted ground-state properties, band structures, and optical responses of rare-earth perovskites~\cite{26,27,28}. The addition of GGA+$U$ corrections is essential to capture the localized character of $f$-electrons and to improve the agreement with experimental observations. Despite this progress, the detailed characterization of cubic RGaO$_3$ perovskites, particularly their electronic and magnetic properties, remain comparatively less explored. A systematic first-principles investigation of these materials is highly desirable, due to large magnetic moments, structural adaptability, and technologically relevant band gaps. Such a study not only provides an insight into the intrinsic properties of rare-earth gallates but also informs the design of advanced materials for spintronic devices, radiation detectors, and high-performance optical components. In this work, we employ DFT within the FP-LAPW framework to investigate the structural, electronic, optical, and magnetic properties of cubic RGaO$_3$ perovskites, where R = Ho, Tm, and Er. By examining lattice parameters, density of states, magnetic moments, and optical spectra, we aim to elucidate how the rare-earth element modulates the physical behavior of these compounds. The results presented herein are expected to complement the existing experimental studies and contribute to a deeper understanding of the complex interactions governing rare-earth perovskite systems.

\section{Computational details}
Density functional theory is implemented to investigate the structural, electronic, optical and magnetic properties of cubic RGaO$_3$ (R = Gd, Tb, Dy) compounds. Spin-polarized investigation is performed by applying full potential linearized augmented plane wave (FP-LAPW)~\cite{29} method as implemented in the WIEN2k code~\cite{30}, designed by Peter and Schwarz~\cite{31}. All calculations are performed under Perdew, Burke and Ernzerhof pseudopotential~\cite{32} implemented within the generalized gradient approximation (GGA) to treat exchange-correlation potential. The GGA and local density approximation (LDA) functional often fail to treat strongly correlated systems to accurately describe $f$-orbitals systems. To address this limitation, GGA+$U$ ($\textrm{U} = 6$~eV) technique is used to treat $f$-electron system~\cite{33}. The choice of $\textrm{U} = 6$~eV was based on systematic testing of different U values to ensure a physically consistent description of the electronic structure and magnetic ground state. This value provided stable spin-polarized solutions and realistic exchange splitting of the localized $4f$ orbitals. The same U value was used for all three compounds because these neighboring lanthanides have similar localized $4f$ states in comparable chemical environments, ensuring a consistent comparison throughout the series. In the FP-LAPW method, the crystal is divided into non-overlapping muffin-tin spheres surrounding each atom and an interstitial region between them. Inside the muffin-tin spheres, wavefunctions are expanded in spherical harmonics, while plane waves are used in the interstitial region. The parameter $l_\textrm{max} =10$ represents the maximum angular momentum quantum number used in the spherical harmonic expansion of the wavefunctions inside the muffin-tin spheres. This relatively high value ensures accurate representation of localized rare-earth $4f$ states, which require higher angular momentum components for precise description. The muffin-tin radii $R_{\textrm{mt}}$ define the atomic sphere size around each atom. These radii were carefully chosen to avoid a sphere overlap while maximizing the sphere volume to improve the basis efficiency and numerical stability. The plane-wave cutoff in the interstitial region is determined by the parameter $K_{\textrm{max}}$, and the product $R_{\textrm{mt}}K_\textrm{max}= 7$ controls the size of the basis set. This value was selected after convergence testing to ensure reliable total energy and charge density convergence without excessive computational cost. The parameter $G_\textrm{max}= 12$ specifies the maximum magnitude of the reciprocal lattice vector used in the Fourier expansion of the charge density. This cutoff ensures a sufficient accuracy in describing charge density variations, particularly important for strongly correlated systems. Brillouin zone integrations were performed using a mesh of 1000 $k$-points to ensure accurate sampling of reciprocal space. The total energy convergence criterion between successive self-consistent field (SCF) cycles was set to 0.001~Ry. The ideal cubic perovskite structure with space group Pm\={3}m was considered. The atomic positions were assigned as A (0,0,0), B (1/2,1/2,1/2), and C (1/2,1/2,0), corresponding to the conventional cubic perovskite arrangement.

\section{Results and discussion}

\subsection{Structural properties}
\begin{figure}[!t]
\centering
\includegraphics[width=0.75\textwidth]{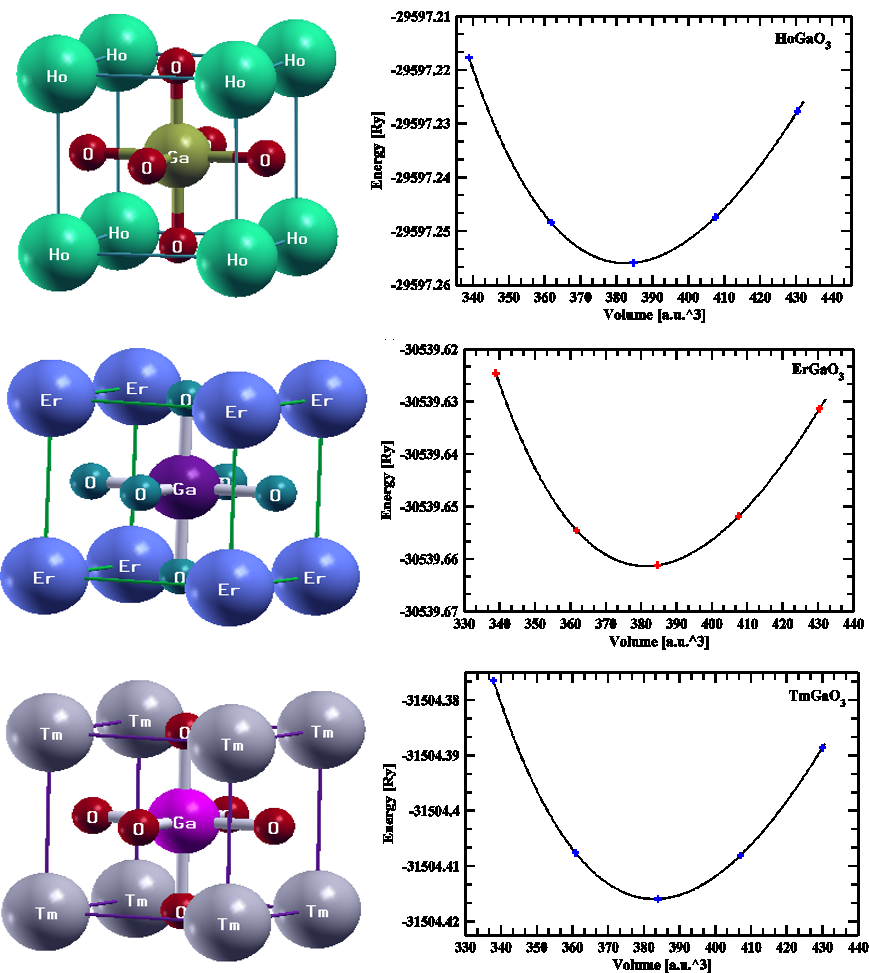}
\caption{(Colour online) Crystal structures and total energy versus volume ($E$--$V$) curves of HoGaO$_3$, ErGaO$_3$, and TmGaO$_3$; the minima indicate the equilibrium volumes.}
\label{fig1}
\end{figure}

To examine the structural properties of the studied perovskite systems, structural optimization was accomplished to know the ground-state parameters, including lattice constants and equilibrium volumes. The general chemical formula of perovskites is ABX$_3$, where A and B are cations and X is an anion. Typically, the A-site cation is larger than the B-site cation. In the ideal cubic perovskite structure (Pm\={3}m symmetry), the A atom inhabits the corners of the cube at (0,0,0), the B atom is positioned at the body center (1/2,1/2,1/2), and the X anion is placed at the face-centered positions (1/2,1/2,0). In this arrangement, the A-site cation exhibits a coordination number of 12, while the B-site cation is octahedrally coordinated by six anions. Further, the tolerance factor serves as an empirical indicator that links the material chemical composition to its structural stability. Thus, it is essential to compute the structural stability of perovskite materials. Goldschmidt defined the tolerance factor of ABO$_3$ perovskite materials by the following relation:

\begin{equation}
t = \frac{(r_{\textrm{A}} + r_{\textrm{B}})}{\sqrt{2(r_{\textrm{B}} + r_{\textrm{O}})}}\,,
\end{equation}
where $r_{\textrm{A}}$, $r_{\textrm{B}}$ and $r_{\textrm{O}}$ are ionic radii of atoms A, B (cations) and O (anion). The calculated tolerance factors for RGaO$_3$ (R = Ho, Er, Tm) are approximately 0.90, placing them within the range typically associated with cubic perovskites. Furthermore, the corresponding octahedral factors for these gallates also fall within the stability window for cubic symmetry, supporting their structural robustness. Rare-earth gallate perovskites RGaO$_3$ (R = Ho, Tm, Er) crystallize in a cubic lattice with space group Pm\={3}m (\#221) as shown in figure~\ref{fig1}. Structural optimizations were carried out using the FP-LAPW method within the PBE-GGA approximation. The total energy was calculated as a function of unit cell volume, yielding the characteristic U-shaped energy--volume curve. The minimum of this curve corresponds to the equilibrium volume ($V_0$) along the $x$-axis and the ground-state energy ($E_0$) along the $y$-axis as shown in figure~\ref{fig1}. Equilibrium structural parameters were obtained by fitting the energy--volume data to the Birch--Murnaghan equation of state (BM-EOS)~\cite{34}. The calculated lattice parameters are summarized in table~\ref{tab1}. Notably, the lattice constants systematically decrease from Ho to Er, consistent with the lanthanide contraction. The optimized crystal structures of the RGaO$_3$ compounds are illustrated in figure~\ref{fig1}. To further evaluate the reliability of the present calculations for the hypothetical cubic HoGaO$_3$, ErGaO$_3$, and TmGaO$_3$ phases, we compared our results with available experimental and theoretical data for rare-earth gallates. Rare-earth orthogallates are well-established perovskite-related compounds~\cite{35}, and high-pressure investigations of La$_{1-x}$ Nd$_x$GaO$_3$ reported bulk moduli of $169-177$~GPa and a pressure-driven tendency toward higher-symmetry structures \cite{36}. These values closely agree with our calculated bulk moduli ($171.74-177.98$ GPa), supporting the reliability of the predicted stiffness and compressibility. Moreover, atomistic simulations of LnGaO$_3$ ($\textrm{Ln = La-Gd}$) effectively reproduced dielectric, elastic, and thermal properties in good agreement with experiment \cite{37}, while first-principles studies of rare-earth gallates also predicted semiconducting electronic structures and rare-earth-driven magnetic features consistent with the trends obtained here \cite{38}. Hence, the good consistency between our results and the known behavior of real rare-earth gallates provides confidence in the physical soundness of the present predictions.

\begin{table}[!b]
\caption{Extracted lattice parameter $a_0$, ground state volume $V_0$, bulk modulus $B_0$ and ground state energy $E_0$ of RGaO$_3$ compounds.\label{tab1}}
\vspace{0.2cm}
\centering
\begin{tabular}{lcccc}
\hline
Compound & $a_0$ (\AA) & $V_0$ (a.u.$^3$) & $B_0$ (GPa) & $E_0$ (Ry) \\
\hline
HoGaO$_3$ & 3.8415 & 382.5725 & 175.21 & $-29597.255975$ \\
ErGaO$_3$ & 3.8376 & 381.3725 & 177.98 & $-30539.661369$ \\
TmGaO$_3$ & 3.8430 & 383.0097 & 171.74 & $-31504.415961$ \\
\hline
\end{tabular}
\end{table}

\subsection{Mechanical properties}

In order to further assess the reliability and practical applicability of the investigated cubic perovskites, the elastic and mechanical properties of RGaO$_3$ (R = Ho, Er, Tm) were systematically examined. Elastic constants provide fundamental insight into the response of a crystal to external mechanical perturbations and are essential for evaluating mechanical stability, stiffness, ductility, compressibility, and anisotropy.

For a cubic crystal, only three independent elastic constants exist, namely $C_{11}$, $C_{12}$, and $C_{44}$. These constants were calculated using the stress--strain method within the FP-LAPW framework. The values obtained are listed in table~\ref{tab2}. Physically, $C_{11}$ represents resistance to longitudinal deformation, $C_{12}$ describes the transverse response under uniaxial stress, and $C_{44}$ corresponds to resistance against shear deformation.

\subsubsection*{Mechanical stability}

The mechanical stability of cubic crystals is governed by the Born stability criteria, which require:

\begin{equation}
C_{11} > 0, \quad C_{44} > 0, \quad C_{11} - C_{12} > 0.
\end{equation}

All calculated elastic constants satisfy these conditions, confirming the mechanical stability of HoGaO$_3$, ErGaO$_3$, and TmGaO$_3$ in the cubic phase.

\subsubsection*{Bulk, shear and Young's moduli}

To obtain macroscopic mechanical parameters, the Voigt–Reuss–Hill (VRH) approximation was employed. The bulk modulus ($B$), shear modulus ($G$), and Young’s modulus ($E$) were calculated using:

\begin{equation}
B = \frac{C_{11} + 2C_{12}}{3},
\end{equation}

\begin{equation}
G = \frac{C_{11} - C_{12} + 3C_{44}}{5},
\end{equation}

\begin{equation}
E = \frac{9BG}{3B + G}.
\end{equation}

The calculated mechanical parameters are summarized in table~\ref{tab3}. The bulk modulus values (170.33, 169.0, and 164.33 GPa for Er, Ho, and Tm compounds, respectively) indicate a considerable resistance to volume compression. The moderate shear modulus values (60.28, 62.40, and 63.15 GPa) reflect reasonable resistance to shear deformation, confirming mechanical rigidity suitable for structural and functional applications. The relatively high Young’s modulus values (162.80, 168.60, and 170.0 GPa) further demonstrate a considerable stiffness and elastic resistance. These results are consistent with the strong interatomic bonding and rigid BO$_6$ octahedral framework characteristic of oxide perovskites.

\subsubsection*{Poisson’s ratio and bonding nature}

The calculated Poisson’s ratio ($\nu$) values (0.335, 0.332, and 0.326) suggest predominantly ionic bonding character. Typically, $\nu \approx 0.25$ indicates ionic bonding, while larger values imply an increased ductility and central-force interactions.

\subsubsection*{Pugh’s ratio and ductility}

The ductility of materials can be assessed using Pugh’s ratio ($B/G$). A critical value of 1.75 separates brittle and ductile behavior. The obtained $B/G$ ratios (2.83, 2.71, and 2.60) exceed this threshold, indicating that all RGaO$_3$ compounds exhibit a ductile behavior.

\subsubsection*{Cauchy pressure}

Cauchy pressure, defined as ($C_{12} - C_{44}$), provides an insight into the bonding character. The positive values (75.95, 72.47, and 70.0 GPa) suggest dominance of non-directional (ionic) bonding and further support the ductile nature of these materials. The positive Cauchy pressure indicates that the cubic perovskites can sustain deformation without brittle fracture while maintaining the lattice integrity.

\subsubsection*{Elastic anisotropy}

The elastic anisotropy factor ($A$) for cubic crystals is defined as:

\begin{equation}
A = \frac{2C_{44}}{C_{11} - C_{12}}.
\end{equation}

For an isotropic material, $A = 1$. The calculated values (0.42, 0.45, 0.47) deviate from unity, indicating en elastic anisotropy in these cubic perovskites.

Overall, the elastic parameters confirm that cubic RGaO$_3$ compounds are mechanically stable, moderately stiff, elastically anisotropic, and predominantly ionic in nature, making them promising candidates for structural, optoelectronic, and spintronic applications.

\begin{table}[!t]
\caption{Calculated elastic constants $C_{ij}$ (GPa) of RGaO$_3$ compounds.\label{tab2}}
\vspace{0.2cm}
\centering
\begin{tabular}{lccc}
\hline
Compound & $C_{11}$ & $C_{12}$ & $C_{44}$ \\
\hline
ErGaO$_3$ & 287.55 & 112.75 & 36.80 \\
HoGaO$_3$ & 285.91 & 111.47 & 39.00 \\
TmGaO$_3$ & 275.30 & 109.20 & 39.20 \\
\hline
\end{tabular}
\end{table}

\begin{table}[!b]
\caption{Calculated mechanical parameters (VRH approximation) of RGaO$_3$ compounds.\label{tab3}}
\vspace{0.2cm}
\centering
\begin{tabular}{lcccccc}
\hline
Compound & $B$ (GPa) & $G$ (GPa) & $E$ (GPa) & $\nu$ & $B/G$ & $A$ \\
\hline
ErGaO$_3$ & 170.33 & 60.28 & 162.80 & 0.335 & 2.83 & 0.42 \\
HoGaO$_3$ & 169.00  & 62.40 & 168.60 & 0.332 & 2.71 & 0.45 \\
TmGaO$_3$ & 164.33  & 63.15 & 170.00 & 0.326 & 2.60 & 0.47 \\
\hline
\end{tabular}
\end{table}

\subsection{Electronic properties}

The electronic behavior of the RGaO$_3$ (R = Ho, Er, Tm) perovskites was examined by calculating their band structures and density of states (DOS) using spin-polarized calculations within the GGA+$U$ formalism ($U=6$~eV). This approach is used for the strong on-site Coulomb interactions in the localized $4f$ orbitals, which are essential for accurately describing the electronic structure of rare-earth-based materials.

\begin{figure}[!t]
\centering
\includegraphics[width=0.7\columnwidth]{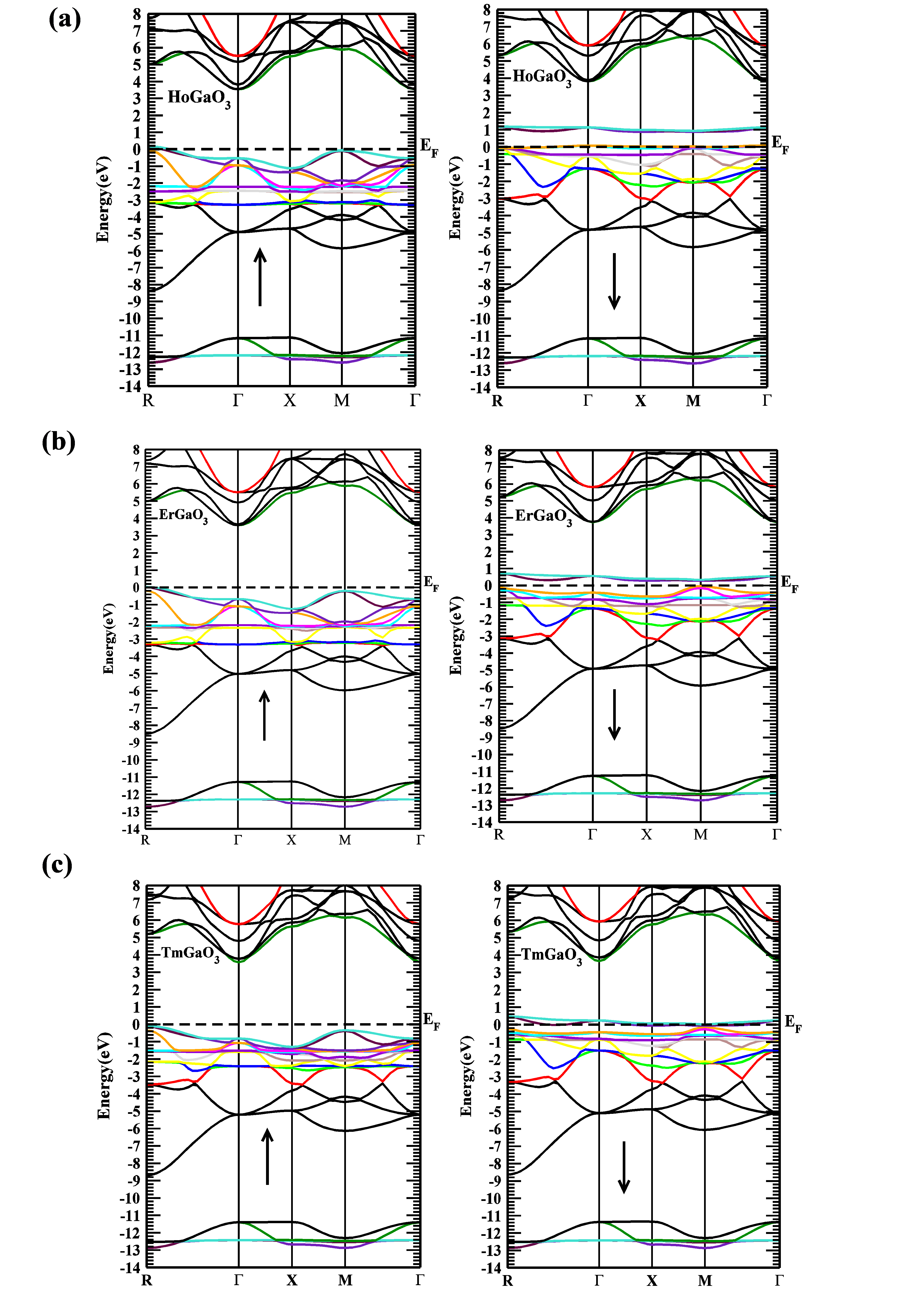}
\caption{(Colour online) Spin-resolved band structures of (a) HoGaO$_3$ (b) ErGaO$_3$ (c) TmGaO$_3$, showing spin-up ($\uparrow$) and spin-down ($\downarrow$) channels.}
\label{fig2}
\end{figure}

\subsubsection{Band structure}
The band structure determines whether a material exhibits metallic, semiconducting, half-metallic, or insulating character, depending on the dispersion and occupation of the valence and conduction bands. The nature of the bandgap-direct or indirect-further defines its optoelectronic properties. In a direct bandgap, the maximum of the valence band and the minimum of the conduction band occur at the same $k$-point in the Brillouin zone, while in an indirect bandgap they are located at different symmetry points. For HoGaO$_3$ figure~\ref{fig2}a, the spin-up channel exhibits an indirect bandgap of 3.4 eV between the R and $\Gamma$ points. By contrast, the spin-down channel shows a metallic behavior with no bandgap. Both semiconducting and metallic spin channels are indicative of half-metallicity, making HoGaO$_3$ potentially useful in spintronic applications. The band structure of ErGaO$_3$ figure~\ref{fig2}b reveals an indirect bandgap of 3.5 eV in the spin-up configuration, also between R and $\Gamma$ points. In the spin-down channel, a small direct bandgap of 0.2 eV is observed, suggesting a semiconducting character in both spin channels, although with a markedly reduced gap for minority spins. For TmGaO$_3$ figure~\ref{fig2}c, the spin-up channel shows an indirect bandgap of 3.6 eV at R--$\Gamma$, while the spin-down configuration is metallic. Similar to HoGaO$_3$, this implies a half-metallic behavior, where the material acts as a conductor for one spin orientation and a semiconductor for the other. Such characteristics are significant for applications requiring high spin polarization, such as magnetic tunnel junctions and spin filters. The pronounced spin-dependent difference in the band gap originates from the strong exchange splitting induced by the localized $4f$ electrons of the rare-earth ions combined with the on-site Coulomb interaction (GGA+$U$). In the spin-up channel, the electronic states are well separated due to exchange polarization, leading to a relatively wide band gap (3.4 eV). However, in the spin-down channel, the exchange interaction shifts the conduction and valence band edges closer together. The partial occupation and hybridization of $4f-2p$ states considerably reduce the gap, resulting in a near-metallic or very narrow-gap semiconducting behavior (0-–0.2 eV). This strong asymmetry between spin channels indicates half-metallic or nearly half-metallic characteristics, which are consistent with the magnetic stabilization obtained at $U=6$~eV. The abrupt gap reduction is therefore a direct consequence of exchange splitting and electron correlation effects captured within the GGA+$U$ framework.

\subsubsection{Density of states}
\begin{figure}[!t]
\centering
\includegraphics[width=0.56\columnwidth]{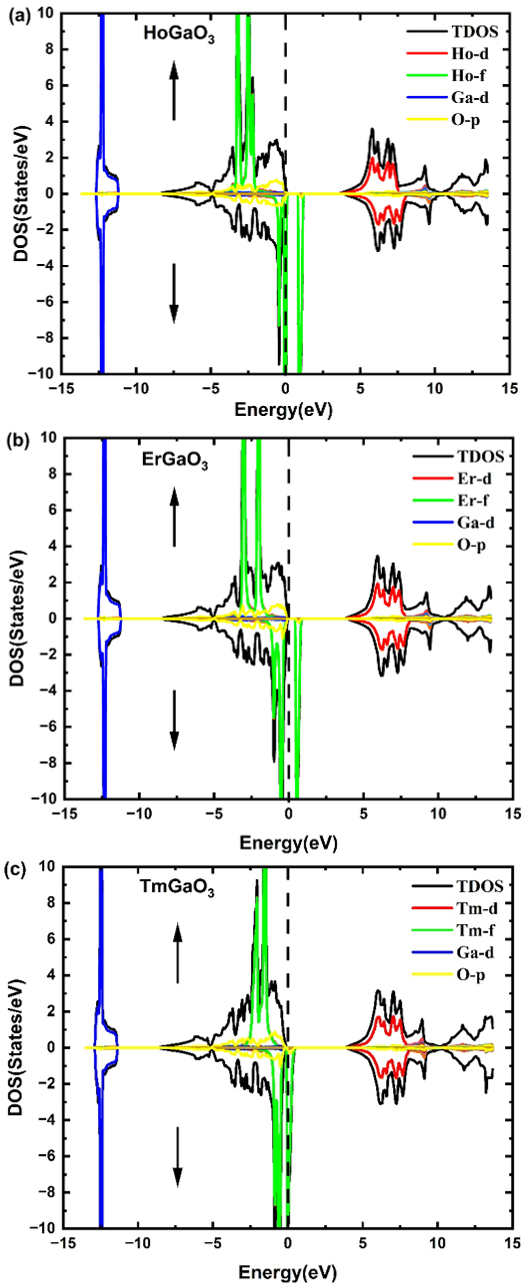}
\caption{Calculated TDOS and PDOS for spin-up ($\uparrow$) and spin-down ($\downarrow$) channels: (a) HoGaO$_3$, (b) ErGaO$_3$, (c) TmGaO$_3$.\label{fig3}}
\end{figure}

The electronic nature of a material can be described more precisely by calculating the total density of states (TDOS) and partial density of states (PDOS). The total density of states (TDOS) and partial density of states (PDOS) of cubic perovskite RGaO$_3$ are calculated by GGA+$U$ procedure given in figure~\ref{fig3}. It is shown in the figure that the whole energy spectrum is divided into two parts, one is beneath the fermi level which is a valence band and the other is above the fermi level which is a conduction band. Density of states of HoGaO$_3$ is shown in figure~\ref{fig3}a. In spin up configuration, Ho-$f$ is leading in the valence band and their energy range is $-4.0 \div (-1.4)$~eV, while O-$p$ state has a very small contribution. In spin down mode, above the fermi level major contribution results from Ho-$f$ state and its energy range is $0.8-1.2$~eV. While, Ho-$d$ state has a small contribution in the conduction band and its energy range is $3.9-8.0$~eV. The total and partial density of states offer insights into the contributions of different atomic orbitals to the electronic structure around the Fermi level. In ErGaO$_3$ figure~\ref{fig3}b, the spin-up DOS reveals that Er-$f$ orbitals dominate the valence band between $-3.8$ and $-1.4$ eV, while Er-$d$ states primarily occupy the conduction band in the $3.8-7.7$~eV range. In the spin-down configuration, Er-$f$ states are again dominant below the Fermi level ($-1.9$ to $-0.1$~eV) and continue to contribute considerably above it ($0.2-0.7$~eV and $3.9-7.9$~eV), indicating a strong localization of $4f$ electrons and their hybridization with $d$-states. For TmGaO$_3$ figure~\ref{fig3}c, the spin-up DOS shows a predominant contribution from Tm-$f$ orbitals in the range $-2.9$ to $-0.6$ eV, while Tm-$d$ states dominate above the Fermi level between 3.8 and 7.9 eV. In the spin-down configuration, Tm-$f$ states remain dominant below the Fermi level ($-1.8$ to 0.5~eV), accompanied by a minor contribution from $d$-states extending up to 7.9 eV. This distribution underscores the key role of the rare-earth $f$-electrons in defining the electronic and magnetic properties of these perovskites.

\subsection{Optical properties}
\begin{figure*}[!t]
\centering
\includegraphics[width=0.9\textwidth]{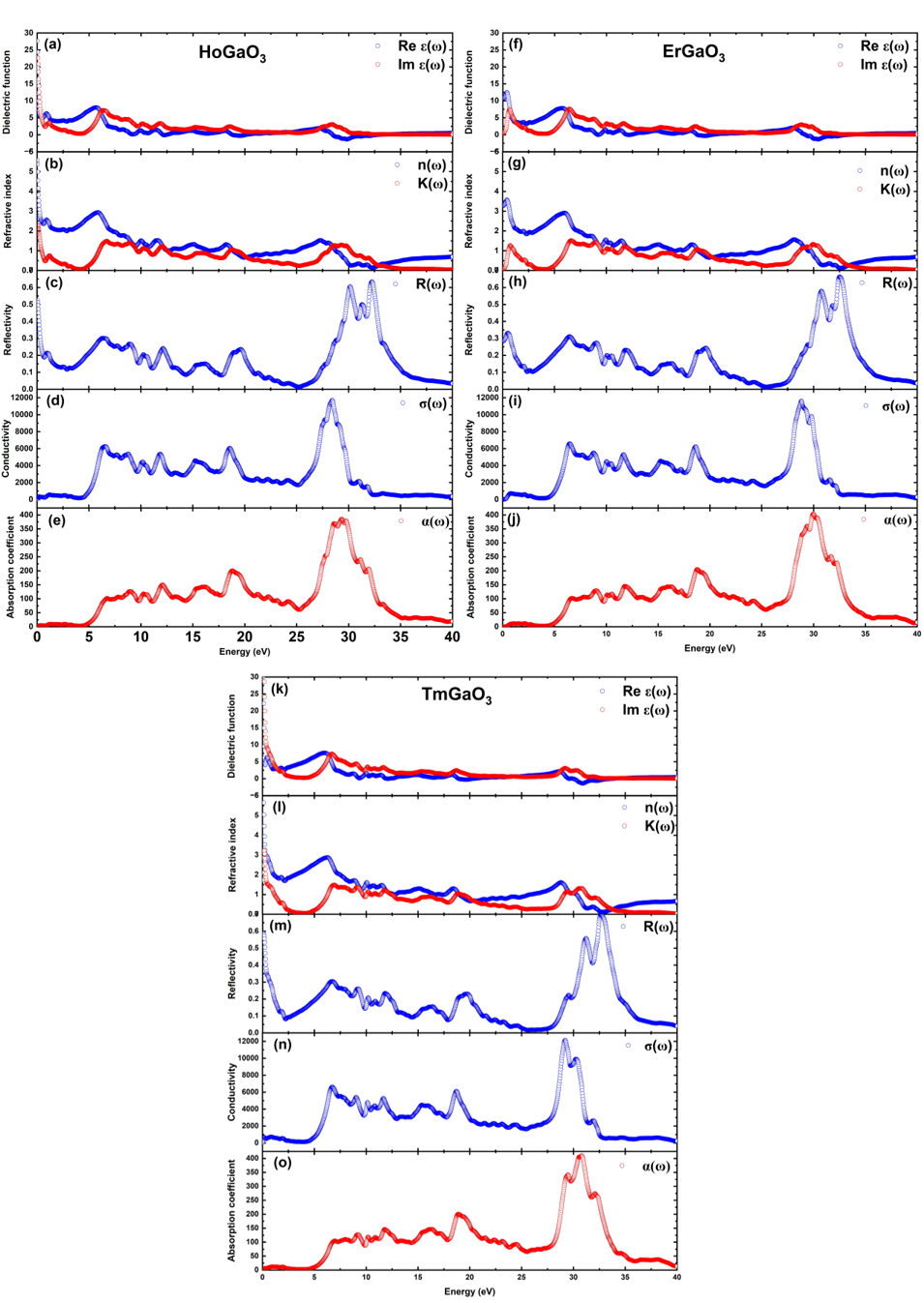}
\caption{(Colour online) Calculated optical properties of RGaO$_3$ (R = Ho, Er, Tm) of real and imaginary parts of the dielectric function; refractive index; absorption coefficient; reflectivity; optical conductivity; and absorption coefficient as a function of photon energy. (a) HoGaO$_3$, (b) ErGaO$_3$, (c) TmGaO$_3$.\label{fig4}}
\end{figure*}

The optical behavior of cubic RGaO$_3$ perovskites (R = Ho, Er, Tm) was investigated by computing the complex dielectric function, refractive index, extinction coefficient, reflectivity, optical conductivity, and absorption coefficient over the photon energy range of $0-42$~eV. These properties are critical for understanding potential applications in optoelectronics, photodetectors, and energy harvesting devices. The frequency-dependent dielectric function is expressed as:

\begin{equation}
\varepsilon(\omega) = \varepsilon_r(\omega) + \ri\varepsilon_i(\omega),
\end{equation}
where $\varepsilon_r$ and $\varepsilon_i$ are the real and imaginary parts, respectively. The imaginary component defines interband electronic transitions and is closely linked to the electronic band structure. The calculated dielectric functions are presented in figure~\ref{fig4}(a, f, k). The real part $\varepsilon_r$ shows pronounced variations at low energies and becomes negative in the $9-19$~eV range, then gradually approaches zero at higher energies. Notably, HoGaO$_3$ and TmGaO$_3$ exhibit negative values near 12~eV, while ErGaO$_3$ shows this behavior at $\sim$9~eV. These negative regions correspond to strong resonances, where plasmonic-like collective oscillations of the valence electrons may occur. The static dielectric constants ($\varepsilon_r$ at $\omega=0$) are summarized in table~\ref{tab4}. The imaginary part $\varepsilon_i$ remains positive throughout the energy continuum, displaying distinct peaks at low photon energies, which reflects a strong optical absorption due to direct interband transitions from occupied valence bands to unoccupied conduction bands. The low-energy onset of $\varepsilon_i$ is consistent with the narrow energy gaps determined from the calculations of the electronic structure.

\subsubsection{Refractive index and extinction coefficient}
Figure~\ref{fig4}(b, g, l) shows the refractive index spectra. All compounds display prominent peaks in the low-energy regime, with ErGaO$_3$ exhibiting the highest refractive index in the ultraviolet range, while HoGaO$_3$ and TmGaO$_3$ have their maxima in the infrared region. This behavior suggests that ErGaO$_3$ may be particularly promising for ultraviolet photonic applications. The extinction coefficient spectra follow similar trends, with peaks corresponding to strong absorption regions. Notably, ErGaO$_3$ show maxima near 6.5--6.7 eV, after which the extinction coefficient diminishes to negligible values at higher energies, representing a reduced absorption in the deep ultraviolet range.

\subsubsection{Reflectivity}
Reflectivity profiles figure~\ref{fig4}(c,h,m) further validate these observations. The optical reflectivity $R(\omega)$ exhibits pronounced features in the low-energy region, consistent with strong interband transitions. Maximum reflectivity values and their associated energies are detailed in table~\ref{tab4}, highlighting the potential of these materials as reflective coatings or mirrors for selective wavelengths.

\subsubsection{Optical conductivity}
The real part of the optical conductivity, displayed in figure~\ref{fig4}(d, i, n), reveals that the threshold energy for the onset of a considerable optical conductivity lies near 5~eV for all investigated compounds. This threshold corresponds to the minimal energy required for electronic excitations across the bandgap. The conductivity peaks occur at energies specified in table~\ref{tab4}, which align with the observed maxima in absorption and extinction.

\subsubsection{Absorption coefficient}
The absorption coefficient spectra are plotted in figure~\ref{fig4}(e, j, o). These curves follow the same general trends as the optical conductivity: absorption begins above $\sim$5~eV and is followed by pronounced maxima at higher photon energies, indicating an efficient photon harvesting in the ultraviolet (UV) range. The energies corresponding to the maximum absorption are summarized in table~\ref{tab4}, which highlights the potential of these perovskites for UV photodetectors and related energy-harvesting applications. Importantly, for half-metallic members (HoGaO$_3$, TmGaO$_3$), the optical response is inherently spin-selective and is governed by their spin-polarized electronic structure. In a half-metal, one spin channel is metallic while the other retains a finite band gap, leading to spin-dependent optical activity: the gapped channel (e.g., $E_g\sim3$~eV) primarily contributes to UV interband absorption, whereas the metallic channel contributes to low-energy (infrared/visible) response via intraband conduction. Consequently, the same compound can exhibit UV absorption associated with the semiconducting spin channel and infrared/low-energy optical response from the metallic channel. This dual-channel behavior is reflected in the dielectric function and reflectivity, where distinct peaks and abrupt variations can be assigned to optical transitions within a specific spin channel. In general, the strong correlation between optical constants and the underlying electronic structure-particularly the rare-earth $f$-Ga $d$ hybridization and its spin dependence-underscores the tunability of optical responses in RGaO$_3$ perovskites and suggests opportunities for spin-selective optical excitation in spintronics/optoelectronic devices.       

\begin{table}[!htbp]
\caption{Optical properties of RGaO$_3$ (R = Ho, Er, Tm) compounds.\label{tab4}}
\vspace{0.2cm}
\centering
\resizebox{\linewidth}{!}{%
\begin{tabular}{lccccccc}
\hline
Compound & $\varepsilon_r(0)$ & $E_n(\omega)$ (eV) & $E_R(\omega)$ (eV) & $R(\omega)$ & $R(\omega)$ (\%) & $E_\sigma(\omega)$ (eV) & $E_\alpha(\omega)$ (eV) \\
\hline
HoGaO$_3$ & 27.6 & 0.01 & 32.2 & 0.63 & 63 & 28.4 & 29.3 \\
ErGaO$_3$ & 10.6 & 3.53 & 32.5 & 0.66 & 66 & 28.8 & 29.4 \\
TmGaO$_3$ & 46   & 0.01 & 32.6 & 0.69 & 69 & 29.1 & 30.8 \\
\hline
\end{tabular}%
}
\end{table}

\subsection{Magnetic properties}
To evaluate the magnetic behavior of RGaO$_3$ (R = Ho, Er, Tm) compounds, spin-polarized density functional theory calculations were performed. The resulting magnetic moments offer an insight into the intrinsic ferromagnetic tendencies of these materials arising from unpaired electrons in the valence shells of the rare-earth elements. The total magnetic moments obtained for each compound are summarized in table~\ref{tab5}. The results reveal that all three compounds exhibit substantial net magnetic moments, consistent with ferromagnetic ordering. Among the atomic constituents, the rare-earth ions (Ho, Er, Tm) contribute the dominant share of the total moment due to their partially filled $4f$ shells, which are known to give rise to localized magnetic moments. For a moment, in HoGaO$_3$, the Ho atom accounts for the majority of the magnetic moment, while oxygen atoms contribute a small positive moment due to spin polarization effects. By contrast, the Ga atom exhibits a minor antiparallel (negative) moment, which slightly offsets the total ferromagnetic alignment, a behavior often attributed to hybridization between Ga $4s/4p$ states and the magnetic orbitals of the rare-earth cation. Additionally, the interstitial regions between atoms were found to contribute non-negligibly to the total moment, reflecting the delocalization of spin density into the lattice voids. Interestingly, the calculated total magnetic moments decrease progressively from HoGaO$_3$ to TmGaO$_3$, following the expected trend with decreasing unpaired $4f$ electrons across the rare-earth series. This systematic reduction in the moment highlights the sensitivity of the magnetic properties to the choice of rare-earth cation and underscores the tunability of these perovskites for spintronic and magneto-optical applications.

\begin{table}[!htbp]
\caption{Net magnetic moment of RGaO$_3$ (R = Ho, Er, Tm) compounds.\label{tab5}}
\vspace{0.2cm}
\centering
\begin{tabular}{lccccc}
\hline
Compound & $M_{\textrm{tot}}$ & $M_{\textrm{int}}$ & $M_{\textrm{R}}$ & $M_{\textrm{Ga}}$ & $M_{\textrm{O}}$ \\
\hline
HoGaO$_3$ & 3.90046 & 0.01094 & 3.76966 & $-0.00122$ & 0.04036 \\
ErGaO$_3$ & 3.00074 & 0.01843 & 2.75367 & $-0.00053$ & 0.07639 \\
TmGaO$_3$ & 2.00921 & 0.02934 & 1.66699 & 0.00231 & 0.10352 \\
\hline
\end{tabular}
\end{table}

\section{Conclusions}
Using density functional theory, we comprehensively examined the structural, electronic, optical, and magnetic properties of cubic RGaO$_3$ (R = Ho, Er, Tm) perovskites via the FP-LAPW approach implemented in WIEN2k. The GGA+$U$ method (U = 6 eV) was applied to accurately capture the electron correlation effects. Structural optimization yielded lattice constants in close agreement with expected trends. Spin-polarized electronic calculations revealed that ErGaO$_3$ is semiconducting, whereas HoGaO$_3$ and TmGaO$_3$ exhibit a half-metallic nature. Optical response of half-metallic compounds (HoGaO$_3$,TmGaO$_3$)  is strongly governed by their spin polarization electronic structure. The semiconducting spin channel absorbs ultraviolet photons, while metallic or small gap channel interacts with infrared light. This dual behavior enables a spin selective optical excitation, relevant for spintronics and optoelectronics applications. The maximum reflectance of TmGaO$_3$ reached 69\%. High-energy optical conductivity further supports their applicability in energy-related devices. Finally, considerable magnetic moments confirm their intrinsic ferromagnetic properties, underscoring their multifunctional potential in spintronic and optoelectronic applications.

\section*{Acknowledgements}
The authors extend their sincere gratitude to Sajjad Qureshi and Ghulam Murtaza for his invaluable technical support, insightful suggestions, and constructive discussions throughout this work. We are deeply grateful to the Higher Education Commission (HEC) for their institutional support, which made this research possible. This study forms part of a thesis submitted to the HEC in partial fulfillment of its academic requirements; however, the findings presented here have not been published elsewhere.


%
%
%

\ukrainianpart
\title[]
{Адаптація структурних, електронних, оптичних та магнітних властивостей галатів рідкісноземельних елементів RGaO$_3$ (R~= Ho, Er, Tm) за допомогою досліджень з перших принципів}

\author
{Т. Усман\refaddr{label1}, 
	К. Ліакат\refaddr{label2}, С.~Хан\refaddr{label3}, M.~Яр~Хан\refaddr{label1}, С.~Алі~Хан\refaddr{label4}}

\addresses{
	\addr{label1} Фізичний факультет Технологічного інституту Цілу, Цзінань 250200, Шаньдун, Китай
	\addr{label2} Науково-технічний університет Кохат (KUST), Кохат, Пакистан
	\addr{label3} Фізичний факультет Науково-технічного університету Банну, Банну 28100, Пакистан
	\addr{label4} Школа матеріалознавства та інженерії, Шаньдунський університет, Цзінань 25000, Китай
}

\makeukrtitle
\begin{abstract}
	Проведено розрахунки ab initio для кубiчних перовскiтiв RGaO$_3$ (R = Ho, Er, Tm) з використанням методу повнопотенцiальної лiнеаризованої доповненої плоскої хвилi (FP-LAPW) в рамках теорiї функцiоналу густини (DFT). Дослiджено їхнi структурнi, електроннi, магнiтнi та оптичнi властивостi. У випадку напівпровідників оптимiзовані сталi ґратки мiнялися вiд 3.83 до 3.85~\AA, тоді як сполуки HoGaO$_3$ і TmGaO$_3$ демонструють напiвметалiчну поведiнку, ймовiрно, внаслiдок спiн-орбiтального зв’язку з електронними станами поблизу рiвня Фермi. Оптичний вiдгук напiвметалiчних сполук (HoGaO$_3$, TmGaO$_3$) визначається їх спiн-поляризацiйною електронною структурою. Напiвпровiдниковий спiновий канал поглинає ультрафiолетовi фотони, тодi як металiчний або малощiлинний канали взаємодiють з iнфрачервоним свiтлом. Така подвiйна поведiнка забезпечує спiн-селективне оптичне збудження, що є важливим для застосувань у спiнтронiцi та оптоелектронiцi. Цi сполуки демонструють високу поверхневу вiдбивну здатнiсть, яка у випадку TmGaO$_3$ досягає максимального значення 69\%. Бiльше того, їхня оптична провiднiсть переважно реалiзується при вищих енергiях, що вказує на їх придатнiсть для високоенергетичних оптоелектронних застосувань. Розрахованi магнiтнi моменти свiдчать про стiйкий феромагнiтний характер у цiй серiї сполук.
	\keywords галати рідкісноземельних елементів, напівметали, теорія функціоналу густини, оптоелектронні властивості, магнітні моменти, спінтронні пристрої
\end{abstract}
\end{document}